# From Concept to Reality: Additive Manufacturing in Particle Accelerator and Storage Ring R&D at GSI and for FAIR

Chuan Zhang [a, 1, 2, 3], Roland Böhm [1], Eduard Boos [1, 2], Ramy Cherif [1], Alexander Japs [1], Stefan Wunderlich [1, 2]

[1] GSI Helmholtz Centre for Heavy Ion Research, Germany
[2] Institute for Applied Physics, Goethe-University Frankfurt, Germany
[3] Helmholtz Research Academy Hesse for FAIR, Germany

**Abstract.** State-of-the-art additive manufacturing technologies are not only finding ever-wider applications in everyday life, but also assuming an increasingly important role in scientific research. This kind of advanced manufacturing method eliminates many of the constraints of conventional processes in fabricating components with complex external shapes or intricate internal structures, thereby providing enhanced flexibility for the design and realization of a new generation of more efficient particle accelerators and storage rings. The RACERS team initiated by the Stochastic Cooling Group at GSI, Germany, is worldwide one of the first teams working on this topic. Based on the metal 3D-printing technology, two novel accelerating structures and one efficient cooling plate for a future stochastic cooling system are under development at GSI and for the FAIR project, respectively. Some successful experience as well as learnt lessons will be presented.

## 1. Introduction

Additive Manufacturing (AM) technologies are being investigated and applied by the RACERS (Research on Additive Construction for Efficient Resonating Structures) team at GSI in Germany for the R&D of next-generation particle accelerators and storage rings. The RACERS team was originally named RACE (Resonators Additively Constructed for Experiments) after its flagship project: the realization of an innovative 704.4 MHz CH (Cross-bar H-mode) cavity using an AM process—metal 3D printing. Following years of research on a new radio-frequency jump method for shortening large accelerators significantly, the proposal and initial design of the novel 704.4 MHz CH structure was published in 2021 [1, 2]. Operating at nearly twice the highest frequency of all previously constructed CH structures (360 MHz), the 704.4 MHz RACE cavity features exceptionally compact dimensions—22 cm in diameter and 33.7 cm in length, comparable to the size of a football (see Fig. 1)—which led to a decision to manufacture it using 3D printing already at the very beginning of the frequency-jump study [1, 2]. The main design parameters of the 704.4 MHz RACE cavity are listed in Table 1.

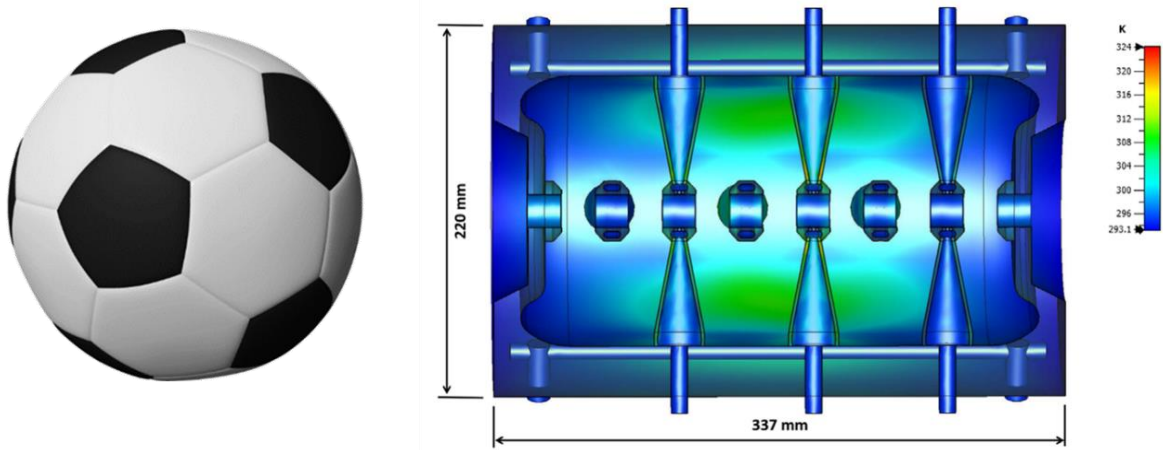

**Fig. 1.** Initial RF design and thermal simulation (2.5 kW RF power as heat load, water-cooled) of the 704.4 MHz RACE cavity based on CST Studio Suite (see [1, 2] for more details).

---

[a] e-mail: `c.zhang@gsi.de`

**Table 1.** Main design parameters of the 704.4 MHz RACE cavity.

| Parameter | Value |
| --- | --- |
| Design particle | proton |
| Frequency $f$ [MHz] | 704.4 |
| Design synchronous beam energy $W_s$ [MeV] | 16.6 |
| $\beta$ (beam velocity relative to the speed of light in vacuum) | 0.186 |
| Number of accelerating cells $N_{cell}$ | 7 |
| Tank external length $L_{tank,external}$ [mm] | 337 |
| Tank inner radius $R_{tank,i}$ [mm] | 80 |
| Tank outer radius $R_{tank,o}$ [mm] | 110 |
| Drift-tube inner aperture radius $R_{tube,i}$ [mm] | 10 |
| Drift-tube outer aperture radius $R_{tube,o}$ [mm] | 18 |

Both initial RF and thermal simulations of the 704.4 MHz RACE cavity [1] were carried out using CST Studio Suite [3]. The RF simulation showed that the RF power consumption for the cavity would be ~1.5 kW. Therefore, in the thermal simulation with the original cooling design [1, 2], an RF power of 2.5 kW—providing a large safety margin—was applied as the heat load to the interior of the cavity, resulting in a maximum surface-temperature rise of ~30°C (see Fig. 1).

In 2022, the mechanical design of the RACE cavity was further developed by introducing the following major modifications, while keeping the parameters summarized in Table 1 unchanged, except $R_{tube,o}$, which was reduced from 18 to 16 mm [4]:

- The tank shape has been changed from cylindrical to octagonal, based on an idea to simplify the assembly.
- The electrodes (each electrode = 1 stem + 1 drift tube + 1 stem) became mountable. As shown in Fig. 2 (a), such an electrode was designed with a smaller base diameter at one end, thereby allowing insertion through the tank wall and subsequent mechanical fixation to the wall.
- All necessary accessories e.g. RF coupler and tuners have been integrated into the design.

As a proof of concept to demonstrate the feasibility of fabricating the 704.4 MHz RACE cavity via the AM technology, a simplified CH-electrode model (highlighted by the red frame in Fig. 2)—consisting of one stem, one drift tube, and a small segment of another stem—was firstly printed with pure copper using a "TruPrint 1000 Green Edition" 3D printer [5, 6].

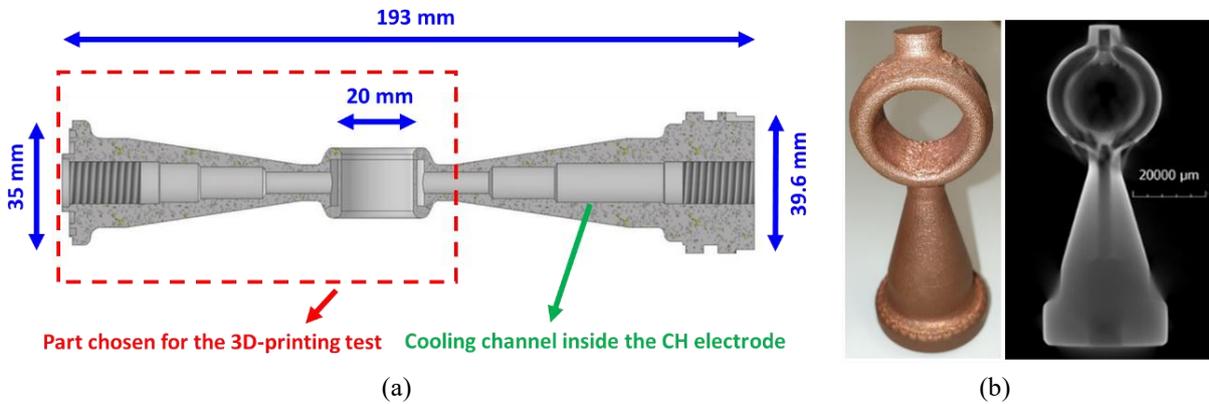

**Fig. 2.** (a) Electrode design for the 704.4 MHz RACE cavity (the cooling-channel design displayed here was later improved by taking advantage of 3D printing); (b) the CH-electrode model printed with pure copper (left) and checked with CT (right) [5, 6].

As one of the main technical approaches for additive manufacturing with metals, the Selective Laser Melting (SLM) method [7] is suitable for fabricating normal-conducting accelerators typically made of copper or stainless steel (afterwards with copper plating). It employs a laser as the energy source to scan the metal-powder bed sequentially—point by point and layer by layer—based on a sliced 3D CAD model. The scanned metal powder will be rapidly melted by the heat of the laser beam and then rapidly solidified. In this way, one can obtain metal parts in any design shape. A key advantage of the SLM method over conventional manufacturing techniques is its ability to overcome many challenges in producing metal components with complex external shapes or intricate internal structures. This is particularly well-suited for constructing the 704.4 MHz RACE cavity, which features small electrodes requiring thin internal cooling channels.

The copper-printed CH-electrode model is shown in Fig. 2 (b). Before post-processing, a "GOM Atos Core 135" 3D scanner was used to assess the surface quality resulting from 3D printing. The measured deviations are mainly within ±75 μm, with most discrepancies occurring near the junctions between the stems and the drift tube. To improve the surface quality, post-processes by means of plasma electrolytic polishing were carried out in cooperation with the Beckmann Institute [8], which reduced the surface roughness from Rz 80 to Rz 6 [5, 6]. To determine the presence of any non-copper constituents in the printed product, X-ray fluorescence analysis was conducted at the GSI Technology Laboratory, revealing a copper content of 99.99 % [5, 6]. As a final step, Computed Tomography (CT), a kind of non-destructive testing technique, was employed to inspect the printed model for internal defects such as pores, blowholes, and cracks. In Fig. 2 (b), the well-printed cooling channels ($Ø_{min}$ = 3.0 mm) inside the model can be clearly seen [5, 6].

## 2. RACE Project: 704.4 MHz CH Cavity

As this cavity is intended for continuous-wave (CW) operation and demands extremely high reliability [1, 2], efficient water-cooling is of critical importance. The successful printing of the CH-electrode model—demonstrating that curved cooling channels with a minimum diameter of 3.0 mm could be reliably printed—motivated the development of a novel cooling design with lotus-root-like channels (see the left graph in Fig. 3) for enhanced thermal performance [5, 6].

Limited by conventional manufacturing, the stem part of a CH electrode is typically cooled by a single thick channel that gradually reduces its diameter in several stages like a telescope and then splits into two thinner channels in the drift-tube section (see the right graph in Fig. 3). The key innovation of the new design is the incorporation of eight $Ø$ = 3 mm slim channels positioned closer to the stem surface to maximize cooling efficiency, which subsequently merge into four channels passing through the drift-tube section—an array of curved, very slim channels unattainable with conventional manufacturing methods.

To evaluate the performance of the proposed lotus-root-like cooling-channel design against the typical telescope-like one, thermal simulations of the CH electrodes were conducted using the professional software ANSYS Discovery [9] (see Fig. 3 and Table 2). In these simulations, all electrodes share the same external dimensions (see Fig. 2 (a)), except for the electrode shown in Fig. 3 (b), which incorporates an additional 1 mm surface allowance as a model for real printing (the reason for incorporating this allowance will be clarified in the subsequent discussion). The remaining settings were determined on the basis of the following considerations:

- The CST simulation described earlier showed that approximately 65 % of the total RF power consumption (2.5 kW was taken with a large safety margin) concentrated on the six CH electrodes, corresponding to an average loss of about 270 W per electrode. Since the electrodes located in the middle of the tank usually have higher RF losses, a heat load of 600 W per electrode (with an even greater safety margin) was adopted for all these thermal simulations.
- The water pressure of 6 bar and the inlet water temperature of 25°C were selected in accordance with the GSI infrastructure specifications.
- In operation, cooling water can induce corrosion and wear on copper components, which shortens the service life of the components. The maximum annual copper loss as a function of water temperature at different water velocities is reported in [10], showing that: (1) the loss rate remains moderate (≤ 0.15 mm/year) for water velocities below 1.20 m/s; (2) in general, at a given velocity, higher water temperatures correspond to larger copper loss.

In Fig. 3, different scenarios have been presented:

- Graph (a) shows a copper CH electrode with novel lotus-root-like cooling channels. The maximum surface temperature of the electrode $T_{surface,max}$ reaches 42.2°C at an inlet water flow rate of $\dot{m}$ = 0.02 kg/s. Correspondingly, the maximum water velocity $v_{water,max}$ is 0.88 m/s and the outlet water temperature $T_{water,o}$ is 32.9°C, respectively. According to [10], the resulting maximum copper loss rate will be approximately 0.08 mm/year. For a minimum wall thickness of 2 mm, this corresponds to an estimated service life of about 25 years. In general, a lifetime exceeding 20 years is already considered satisfactory for accelerators.
- Graph (c) also shows a copper CH electrode, but with typical telescope-like cooling channels. It can be seen that the inlet water flow rate must be increased to 0.07 kg/s to also achieve $T_{surface,max}$ = 42.2°C, but this will result in a fourfold increase in water velocity i.e. $v_{water,max}$ >> 1.20 m/s. To mitigate the risk of a significant reduction in component lifetime, stainless steel 316L, due to its substantially higher hardness than copper, is often chosen as the construction material, as in the MYRRHA project [11]. With stainless steel 316L of considerably lower thermal conductivity, a tenfold increase in inlet water flow only yields a maximum surface temperature of 66.6°C, as shown in Graph (d), still significantly higher than in the copper case of Graph (a). Further increasing the water velocity cannot reduce $T_{surface,max}$ below 60°C.
- Graph (b) presents a modified version of the electrode from Graph (a), which was adopted for the final production. Instead of starting eight cooling channels directly at both ends, two short transition sections were introduced to simplify the water connections, with negligible impact on cooling performance. If the

1 mm surface allowance introduced in this model is removed in the simulation, $T_{surface,max}$ would be even lower.

In short, under the same thermal load of 600 W, the novel cooling-channel design can reduce the maximum electrode surface temperature rise by over 50% and yield a more homogeneous surface temperature distribution (note: the colorbar scales in Fig. 3 are different; otherwise, the surface-temperature distributions in Graphs (a–c) would not be distinguishable).

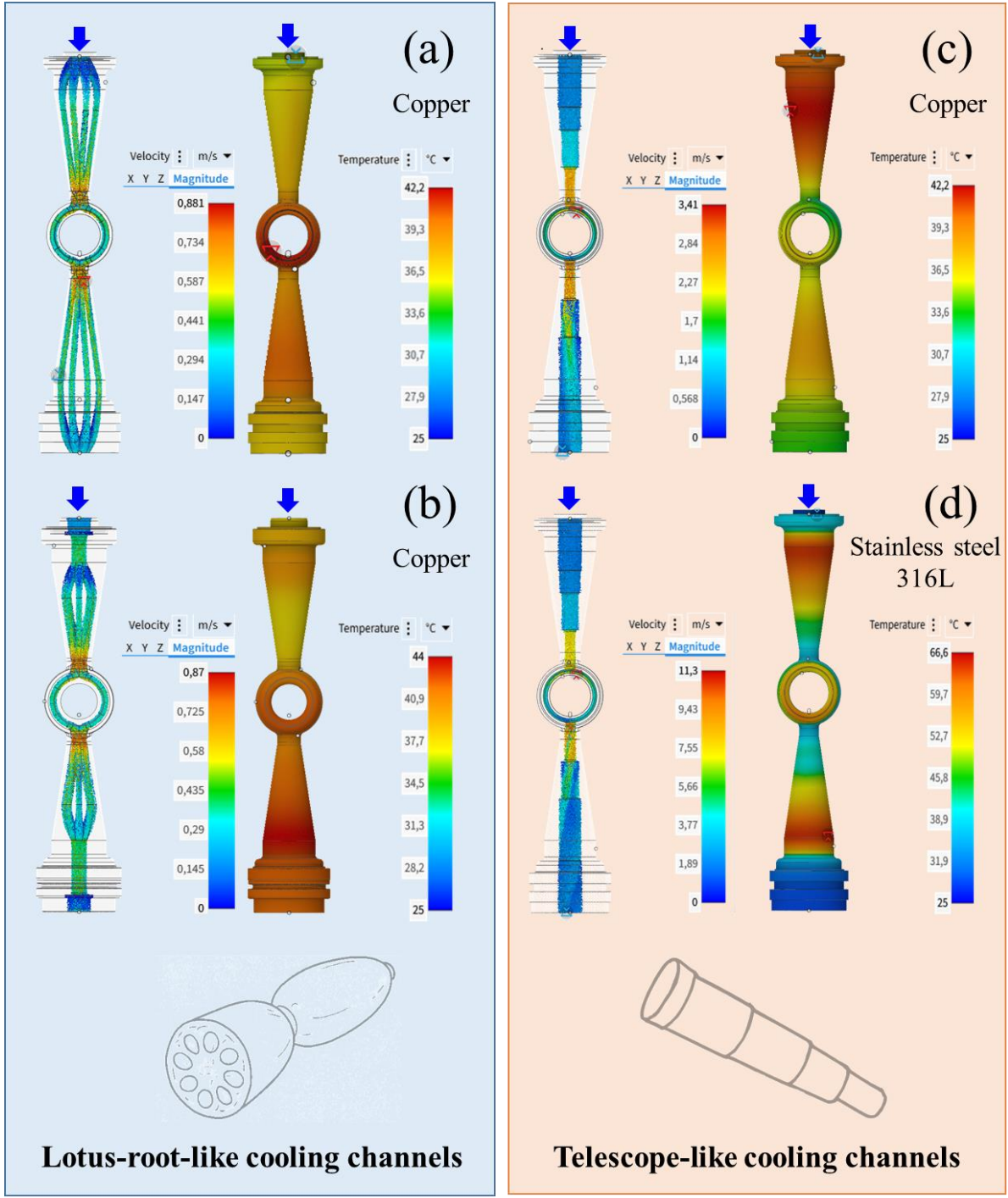

**Fig. 3.** Comparison of novel (left) and typical (right) cooling-channel designs for the 704.4 MHz CH electrodes with ANSYS Discovery simulations. The blue arrows denote the water flow directions.

**Table 2.** Comparison of ANSYS Discovery simulation results of the 704.4 MHz CH electrodes with novel and typical cooling-channels.

| Parameter | Graph (a) | Graph (b) | Graph (c) | Graph (d) |
|---|---|---|---|---|
| Cooling channels | lotus-root-like | lotus-root-like | telescope-like | telescope-like |
| Material | pure copper | pure copper | pure copper | stainless steel 316L |
| Thermal load [W] | 600 | 600 | 600 | 600 |
| Water pressure [bar] | 6 | 6 | 6 | 6 |
| Inlet water temperature $T_{water,i}$ [°C] | 25 | 25 | 25 | 25 |
| Inlet water flow rate $\dot{m}$ [kg/s] | 0.02 | 0.02 | 0.07 | 0.2 |
| Outlet water temperature $T_{water,o}$ [°C] | 32.9 | 32.1 | 27.0 | 25.8 |
| Max. water velocity $v_{water,max}$ [m/s] | 0.88 | 0.87 | 3.41 | 11.30 |
| Max. surface temperature on the electrode $T_{surface,max}$ [°C] | 42.2 | 44.0 | 42.2 | 66.6 |

With the integration of this unique cooling-channel design, the final mechanical design of the 704.4 MHz RACE cavity was completed in 2024 (see Fig. 4 (a)). In the same year, the whole cavity was fabricated using AM, with the electrodes printed in pure copper and the tank in stainless steel 316L (see Fig. 4 (b)). From the printing practice of the CH electrode model, it was learned that with the current 3D-printing technology, the surface quality of printed accelerator components is still insufficient for direct use. Therefore, it was decided to apply an additional 1 mm allowance to all surfaces of the parts during printing, which should be later removed by milling to achieve the final dimensions. Both printed tank and electrodes achieved a surface roughness of Rz ≤ 80. After sandblasting and mechanical milling, the printed tank achieved a surface roughness of Rz 6.3–10, while the electrodes reached an even better finish of Rz < 6.3.

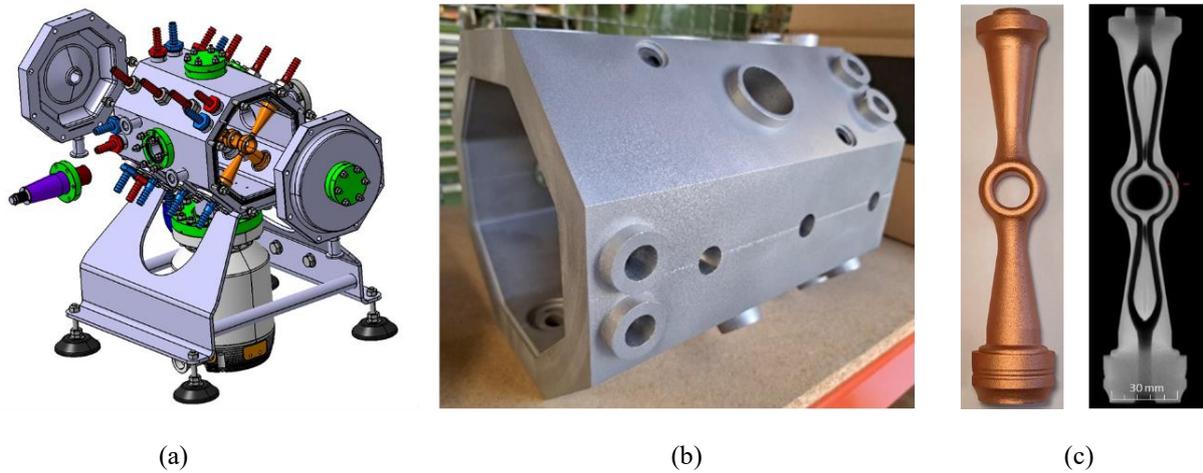

(a)  (b)  (c)

**Fig. 4.** (a) Mechanical design of the 704.4 MHz RACE cavity; (b) the tank printed with stainless steel 316L before post-processing; (c) an electrode printed with copper (left) and scanned with CT (right) before post-processing.

Using a testing chamber equipped with an ionization vacuum gauge and a mass spectrometer (see Fig. 5 (a)), vacuum measurements of a printed CH electrode were carried out in three phases:
- Phase 1: before post-processing (the electrode surface may have rough areas with Rz ≤ 80).
- Phase 2: after post-processing (the electrode has a smooth surface with Rz < 6.3).
- Phase 3: with the two open ends of the electrode sealed by brazing.

For each phase, the experimental procedure comprises the following steps: (1) pump down the chamber and measure the pressure and outgassing rate without bake-out; (2) apply a bake-out, pump down the chamber again, and continue the measurements; (3) repeat Step 2 once or twice. The detailed pressure results are summarized in Table 3. Figure 5 (b) depicts the measured mass spectrum after a cumulative pumping time of 134 h in Phase 3, indicating that the main outgassing species are $H_2O$, $N_2$, CH–N+O, and $CO_2$, all of which remain at very low levels. It can be concluded that the 3D-printed electrode is vacuum compatible, because in general, a vacuum level of $10^{-7}$ mbar is sufficient for linear accelerators.

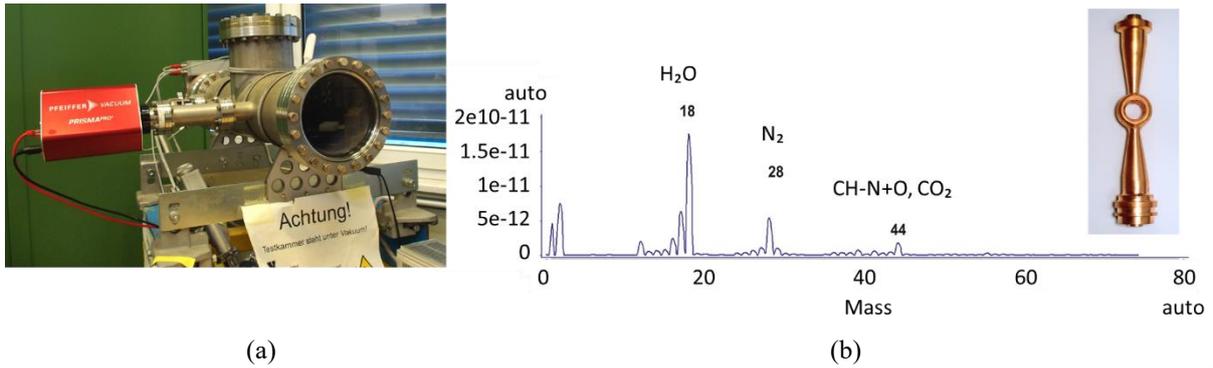

**Fig. 5.** (a) Vacuum test bench for the printed electrode; (b) measured mass spectrum after a cumulative pumping time of 134 h in Phase 3.

**Table 3.** Vacuum test results of a copper-printed 704.4 MHz CH electrode.

| Phase | No bake-out | + 1st bake-out | + 2nd bake-out | + 3rd bake-out |
|---|---|---|---|---|
| 1: before post-processing | $2.1*10^{-7}$ mbar (2 h[#]) | $2.0*10^{-9}$ mbar (45 h) | $1.5*10^{-9}$ mbar (66 h) | -- |
| 2: after post-processing | $1.3*10^{-7}$ mbar (1 h) | $2.3*10^{-9}$ mbar (21 h) | $1.2*10^{-9}$ mbar (45 h) | $1.0*10^{-9}$ mbar (69 h) |
| 3: with the two open ends of the electrode sealed | $5.9*10^{-10}$ mbar (68 h) | $2.3*10^{-10}$ mbar (86 h) | $2.3*10^{-10}$ mbar (110 h) | $2.0*10^{-10}$ mbar (134 h) |

[#]: the cumulative vacuum time is given in brackets.

Following post-processing and vacuum tests, the 704.4 MHz RACE cavity was assembled (see Fig. 6). Subsequently, low-level RF tests of the cavity were conducted at IAP Frankfurt in February 2025 and repeated at GSI in June 2025 (see Fig. 7). The design frequency was achieved, and the measured gap-voltage distribution closely matched the CST simulation results. Currently, the stainless-steel tank is undergoing copper plating. In the near future, vacuum tests, more precise low-level RF measurements, high-power RF tests, and beam experiments will be conducted on the finalized cavity.

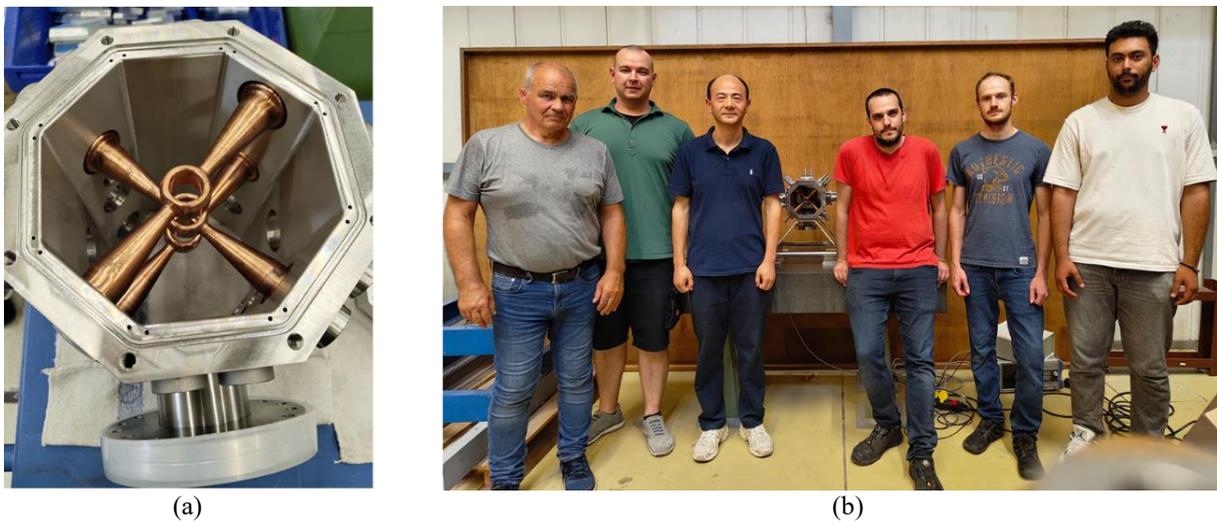

**Fig. 6.** (a) RACE cavity during assembly with printed components after mechanical milling; (b) The RACERS Team with the assembled "table-top" RACE cavity.

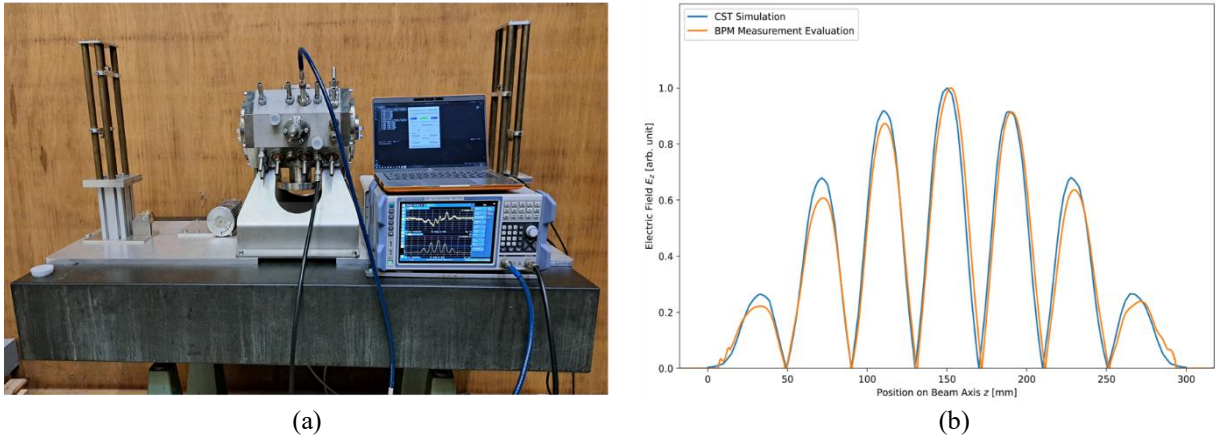

(a) (b)

**Fig. 7.** (a) Test bench for the bead-pull measurements at GSI; (b) simulated and measured gap-voltage flatness along the 704.4 MHz RACE cavity.

## 3. SHAPER Project: 704.4 MHz SH Cavity

During the R&D of the 704.4 MHz RACE cavity, parallel efforts have been also made to explore the feasibility of testing the printed cavity with real beams. For performing a beam test, an existing linac working at 704.4 MHz or its fractional frequencies is needed. A possible choice is to put the 704.4 MHz RACE cavity at the end of the 88.0525 MHz SPIRAL2 linac [12] at GANIL in FRANCE by means of an eightfold frequency jump. For the possible beam test, some boundary conditions given by GANIL are as follows:

- For protons, the output beam energy must be less than 33 MeV, as limited by radiation safety regulations; therefore, an input beam energy $W_{beam}$ = 32 MeV and an energy gain $\Delta W_{beam}$ < 1 MeV have been suggested.
- To avoid particle losses in the beam experiment, the drift-tube inner aperture radius $R_{tube,i}$ should be ≥ 22 mm.

Obviously, the 704.4 MHz RACE cavity designed for a much lower beam energy (16.6 MeV) and a much smaller drift-tube inner aperture radius (10 mm) cannot meet these specified requirements. Therefore, it was decided to develop and print a new cavity. Similar to the 704.4 MHz RACE cavity, the new cavity will have identical accelerating cells working at synchronous phase $\varphi_s$ = -90° (for easy construction), but it should have $R_{tube,i}$ = 22 mm and eight accelerating cells (one more cell due to the eightfold frequency jump) so that the cavity can always see a beam bunch inside during beam operation. In order to achieve beam acceleration with such a $\varphi_s$ = -90° cavity, a special beam dynamics design concept so-called EQUUS (EQUidistant mUltigap Structure) [13] with an asynchronous beam injection (typically the phase and the energy of input beam are less negative and lower than the synchronous phase and energy, respectively) has been adopted. For the new cavity, the design synchronous beam energy is chosen as 33 MeV.

Another remarkable change is that the new cavity will operate in a different mode, Sextupole H-mode (SH) i.e. $H_{31(0)}$ mode. So far, normal-conducting IH (Interdigital H-mode i.e. $H_{11(0)}$ mode) structures and normal- / superconducting CH (Crossbar H-mode i.e. $H_{21(0)}$ mode) structures have been well developed and widely applied [14, 15]. As shown in Fig. 8, the proposed SH structure is a new member for the H-mode DTL family. For this study, a normal-conducting SH as well as the worldwide first SH (to our best knowledge) is being developed, but potentially this structure can be also applied for a superconducting accelerator.

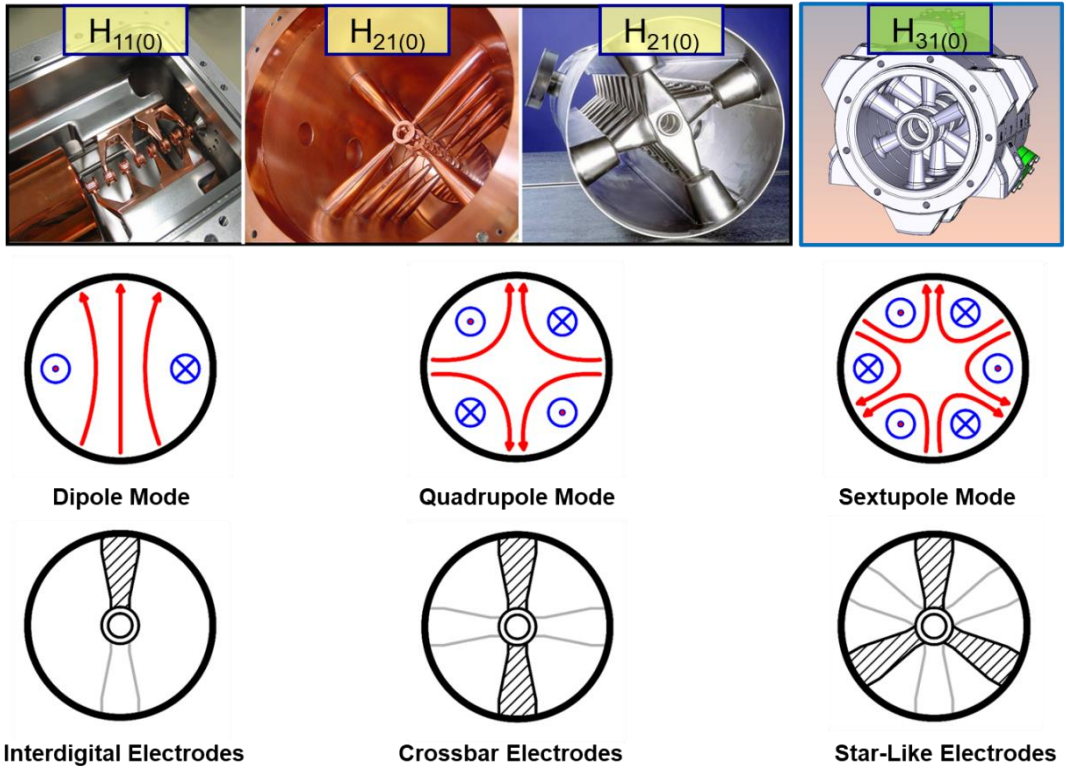

**Fig. 8.** Expansion of the H-mode drift-tube linac family with the new SH structure (top: cavity examples; center: resonating modes; bottom: layout of electrodes).

A comparison of the main design parameters of the newly proposed SH with the RACE CH cavity is given in Table 4. It can be seen that changing the structure from CH to SH offers the following advantages (also see Fig. 9):

- The SH has larger transverse dimensions at the same frequency (more space for cooling channels and accessory components).
- CST Simulations show that the SH has a better shunt impedance value than the CH in case of $R_{tube,i} = 22$ mm.
- The SH has a more favorable RF power loss distribution, with 90% of losses occurring in the stems and 10% in the drift tube (a ratio of 9:1), compared to the CH structure's 75% losses in the stems and 25% in the drift tube (a ratio of 3:1). The SH stems can be cooled more easily.
- Meanwhile, the SH's power loss distribution is more uniform.
- In addition, the SH with star-like electrodes is mechanically more stable.

**Table 4.** Comparison of main design parameters of the 704.4 MHz SH and CH cavities.

| Parameter | 704.4 MHz SH | 704.4 MHz CH |
|---|---|---|
| Resonating mode | $H_{31(0)}$ | $H_{21(0)}$ |
| $N_{cell}$ | 8 | 7 |
| $W_s$ [MeV] | 33 | 16.6 |
| $\beta$ | 0.259 | 0.186 |
| $L_{tank,external}$ [mm] | 495 | 337 |
| $R_{tank,i}$ [mm] | 90.2 | 80 |
| $R_{tank,o}$ [mm] | 112 | 110 |
| $R_{tube,i}$ [mm] | 22 | 10 |
| $R_{tube,o}$ [mm] | 30 | 18 (16) |
| Printing materials | Electrodes & Tank: stainless steel 316L (with copper plating) | Electrodes: copper Tank: stainless steel 316L (with copper plating) |

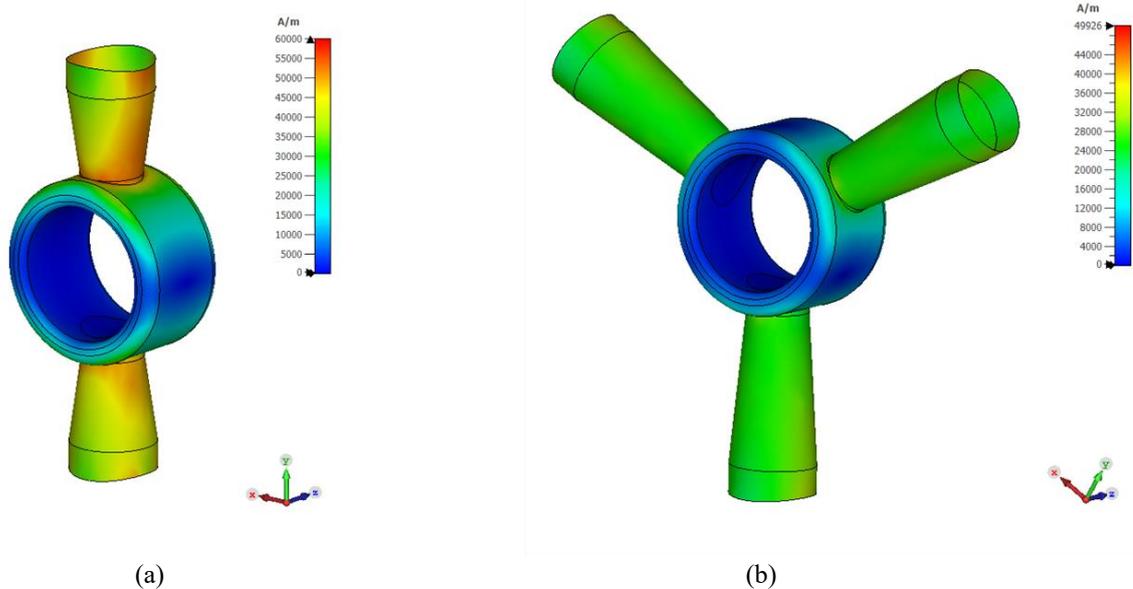

(a) (b)

**Fig. 9.** Comparison of the simulated RF power loss on the electrodes of the CH (left) and SH (right) cavities (for both cases: $f$ = 704.4 MHz, $R_{tube,i}$ = 22 mm)

For the beam dynamics simulation of the SH cavity, a realistic input beam distribution with 73678 macro-particles (see Fig. 10 (a)) was provided by GANIL [16]. For this beam, the beam intensity and energy are $I_{beam}$ = 5 mA (in our simulation, $I_{beam}$ = 40 mA is taken due to the eightfold frequency jump) and $W_{beam}$ = 31.977 MeV, respectively. The output distributions simulated by the LORASR [17] code are shown in Fig. 10 (b). It can be seen that through the SH, the beam quality has been kept good (the emittance growths for both longitudinal and transverse planes are < 2%), and no particle loss has been observed (the beam size is still much smaller than the drift-tube inner aperture). The simulation predicts an energy gain of 311 keV, satisfying the requirement to keep the output beam energy far below 33 MeV. Confirmed by the beam diagnostics team at GANIL, this energy gain can be well measured using the time-of-flight method.

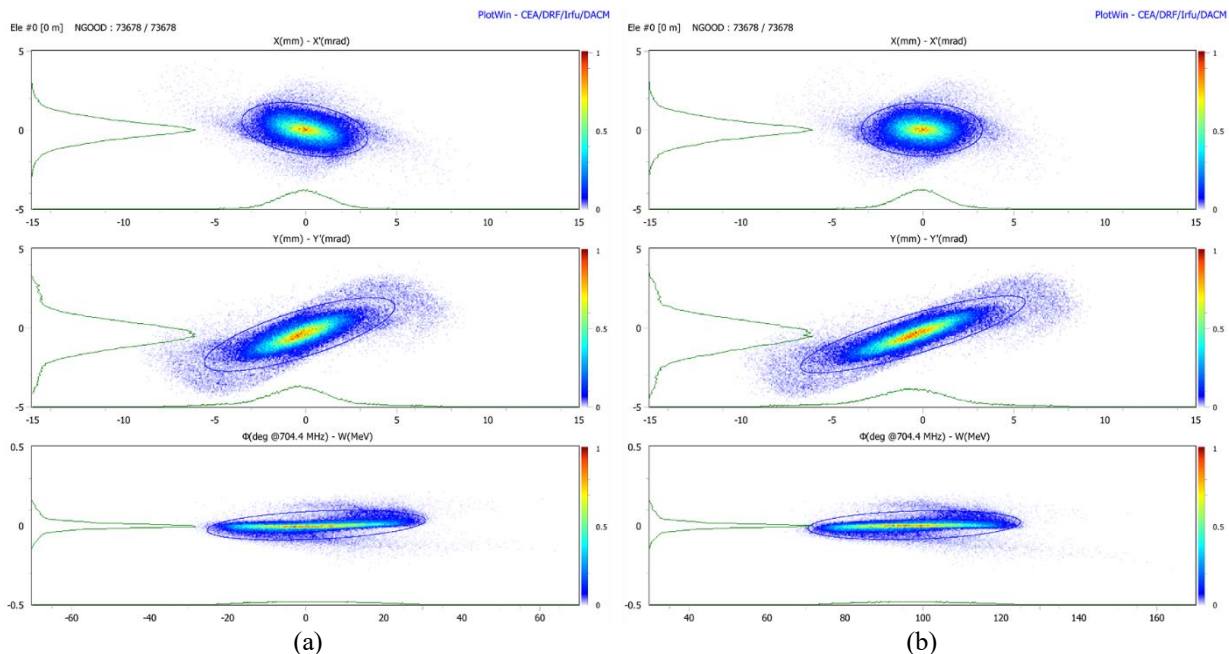

(a) (b)

**Fig. 10.** (a) Input particle distribution from GANIL (with an eightfold frequency jump); (b) output particle distribution simulated with the LORASR code.

The CST simulation showed that the total RF power consumption of this cavity would be 18.1 kW (30.60%, 68.34%, and 1.06% on the tank, electrodes and tuners, respectively) for achieving $\Delta W_{beam}$ = 311 keV. As this SH cavity is very different than the RACE CH cavity, a new project so-called SHAPER (SH Additively Produced as an Efficient Resonator) has been launched.

For the SHAPER cavity, it was decided to firstly test stainless steel 316L as the printing material, since the SH cavity is relatively larger in size and its geometry is relatively easier for cooling. Figure 11 (c) shows the mechanical design of the 704.4 MHz SH cavity and one of the printed SH cavity pieces. After two rounds of sandblasting, the surface roughness of this SH piece reaches an Rz value below 8–10 µm. To the best of our knowledge, the SHAPER cavity is the first sextupole H-mode drift-tube linac structure ever built.

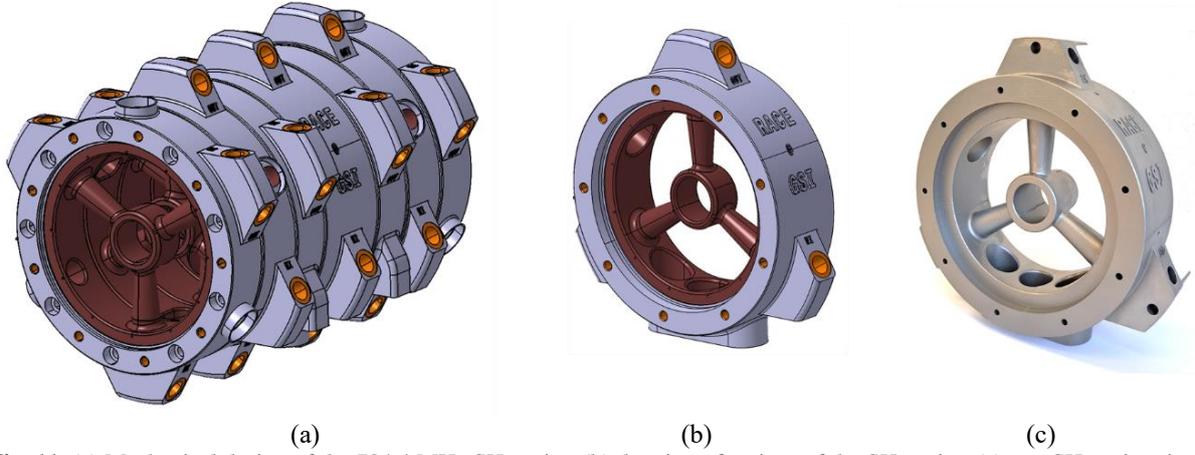

**Fig. 11.** (a) Mechanical design of the 704.4 MHz SH cavity; (b) drawing of a piece of the SH cavity; (c) one SH cavity piece printed with stainless steel 316L.

According to the CST simulation, the average RF power loss per SH electrode is approximately 1.77 kW, with 90% dissipated on the stems and 10% on the drift tube. For the thermal simulations, a heat load of 2.5 kW per electrode, incorporating a large safety margin, was accordingly applied. As illustrated in Fig. 12 (a), each SH electrode is equipped with three independent cooling-channel sets. The cooling-water flow is arranged such that water enters through the outer pipe at the inlet of one stem and exits via the central pipe at the outlet of a neighboring stem, ensuring that the electrode surface is continuously exposed to fresh coolant.

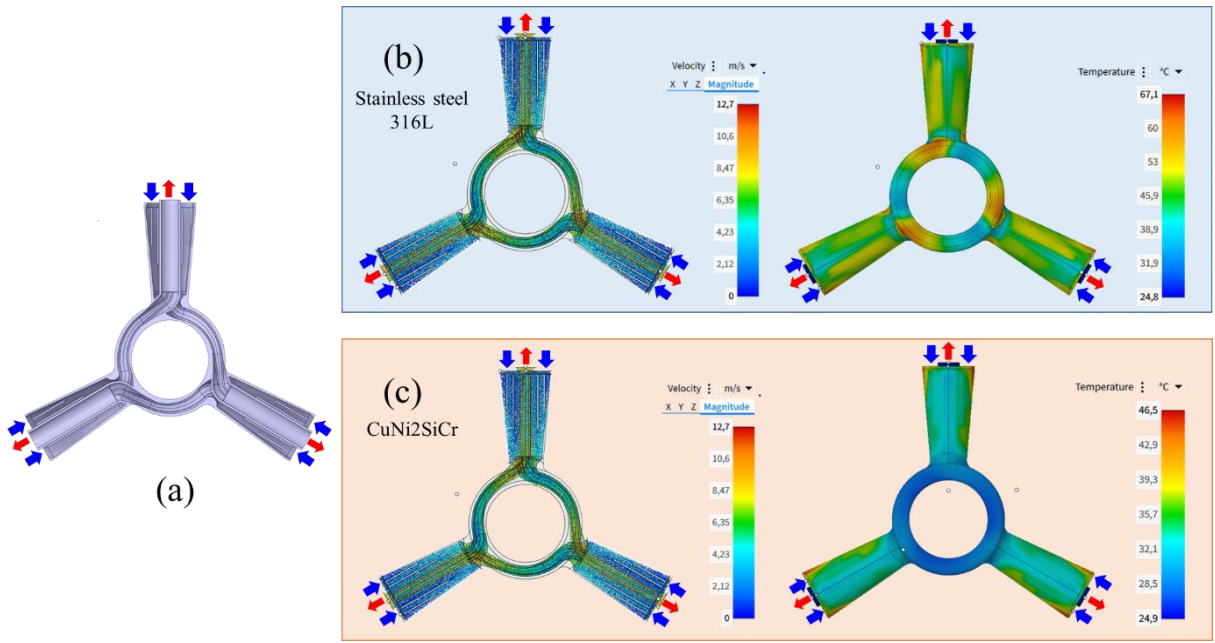

**Fig. 12.** (a) Preliminary cooling-channel design for the SH electrodes; (b) water-velocity (left) and surface temperature (right) distribution simulated by the ANSYS Discovery (material: stainless steel 316L); (c) water-velocity (left) and surface temperature (right) distribution simulated by the ANSYS Discovery (material: CuNi2SiCr). The arrows denote the water flow directions.

Figure 12 (b) shows, an electrode made of stainless steel 316L reaches $T_{\text{surface,max}} = 67.1°C$ with the setting shown in Table 5. To lower $T_{\text{surface,max}}$, one approach is to further optimize the cooling-channel design, while another is to replace stainless steel 316L with a material that offers higher thermal conductivity but comparable hardness.

**Table 5.** Comparison of ANSYS Discovery simulation results of the 704.4 MHz SH electrodes made of stainless steel 316L and CuNi2SiCr.

| Parameter | Stainless steel 316L | CuNi2SiCr [#] |
|---|---|---|
| Thermal load [W] | 2500 | 2500 |
| Water pressure [bar] | 6 | 6 |
| Inlet water temperature $T_{water,i}$ [°C] | 25 | 25 |
| Inlet water flow rate $\dot{m}$ [kg/s] | 0.06 * 3 | 0.06 * 3 |
| Outlet water temperature $T_{water,o}$ [°C] | 25.4 | 25.3 |
| Max. water velocity $v_{water,max}$ [m/s] | 12.7 | 12.7 |
| Max. surface temperature on the electrode $T_{surface,max}$ [°C] | 67.1 | 46.5 |

[#]: The thermal conductivity adopted for this simulation is 190 W/(m·K).

In addition to the conventional construction materials for accelerators, such as copper and stainless steel, we are also exploring the use of high-performance copper alloys, such as CuNi2SiCr and CuCr1Zr, for the additive manufacturing of accelerator components. The key material properties of these copper alloys are compared with those of pure copper and stainless steel 316L in Table 6. It can be seen that CuNi2SiCr can provide hardness comparable to stainless steel 316L while offering much higher thermal conductivity. Figure 12 and Table 5 indicate that replacing stainless steel 316L with CuNi2SiCr as the SH electrode material reduces $T_{surface,max}$ to 46.5°C under the same conditions. It should be noted that the simulation presented in Fig. 12 (c) was performed using the lower-end value of the thermal conductivity range of CuNi2SiCr namely 190 W/(m·K).

**Table 6.** Properties of possible construction materials for accelerators.

| Parameter (at 20°C) | Pure copper [#1] | Stainless steel 316L [#2] | CuNi2SiCr [#3] | CuCr1Zr [#1] (heat-hardened) |
|---|---|---|---|---|
| Thermal conductivity [W/(m·K)] | 397 | 15 | 190-240 | 310 |
| Specific electrical conductivity [MS/m] | 57-59 | 1.55 | 23 | 43 |
| Coefficient of thermal expansion [/°C] | 16.5 ·10⁻⁶ | 16 ·10⁻⁶ | 15·10⁻⁶ | 17 ·10⁻⁶ |
| Specific electrical resistivity [(Ω·mm²)/m] | 0.0175 | 0.75 | 0.0435 | 0.023 |
| Density [g/cm³] | 8.94 | 7.95 | 8.84 | 8.91 |
| Hardness (Vickers hardness, HV) | 45–100 | 150–220 | 160–240 | 110–200 |

[#1]: data by Kupferverband, Germany, https://kupfer.de/
[#2]: data by Assemblean GmbH, Germany, https://assemblean.com/
[#3]: data by Metalcor GmbH, Germany, https://www.metalcor.de/

It is also worth to mention that many R&D efforts for the CH and SH cavities benefit from dual use, e.g.:
- Shared infrastructure, such as the vacuum test bench and bead-pull setup, supports both CH and SH cavities.
- Many accessories, including RF coupler and tuners, can employ same or similar designs for both cavity types.
- A single RF amplifier can be used to test both cavities up to a certain RF power level.

## 3. SCO Project: Efficient Cooling Plate for Stochastic Cooling Kickers

The AM technologies can be valuable for the R&D of not only particle accelerators but also storage rings. The future Collector Ring (CR) of the FAIR facility will need a 1–2 GHz stochastic cooling system for providing a fast 3D cooling of hot secondary beams (antiprotons at 3 GeV and rare isotope ions at 740 MeV/u) at intensities up to $10^8$ particles per cycle [18]. As shown in Fig. 13, the CR Stochastic Cooling (CR-SC) system mainly consists of the following sub-systems:
- Two cryogenic plunging slotline pick-up tanks (one for horizontal and longitudinal cooling; one for vertical and longitudinal cooling).
- One Palmer pick-up tank.
- Two slot-ring kicker tanks (one for transverse cooling; one for longitudinal cooling).
- The RF signal processing chains between the tanks.

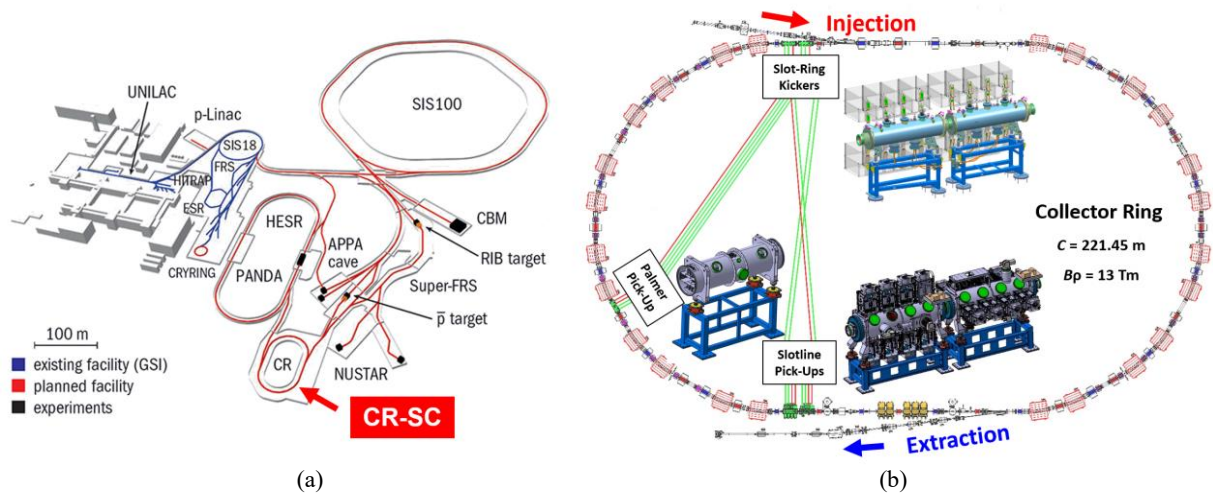

|  |  |
|:-:|:-:|
| (a) | (b) |

**Fig. 13.** (a) The existing GSI facility (blue) and the future FAIR facility (red); (b) the planned CR Stochastic Cooling (CR-SC) system.

The kicker part has been designed by Forschungszentrum Jülich (FZJ), Germany. Each of the two kickers contains 128 slot-rings and every 16 slot-rings are grouped into 1 stack and attached to 8 divider boards (see Fig. 14). ANSYS simulations performed by FZJ [19] showed that when the power dissipated in the resistors of a divider board was increased to 95 W for 5 minutes, (1) the temperature of the resistors rose to approximately 110°C within 0.1 seconds; (2) a subsequent temperature increase, driven by the heating of the entire board, continued until reaching a maximum of nearly 174°C; and (3) the resistors exceeded their maximum permissible operating temperature of 140°C after approximately 100 seconds.

Therefore, the originally employed cooling method—attaching a heat sink to one edge of the divider board (see Fig. 14)—carries the risk of insufficient thermal management. For a more efficient cooling solution, the RACERS team proposed developing a thin cooling plate featuring slim internal channels fabricated via additive manufacturing, which can be attached to the rear surface of the divider board.

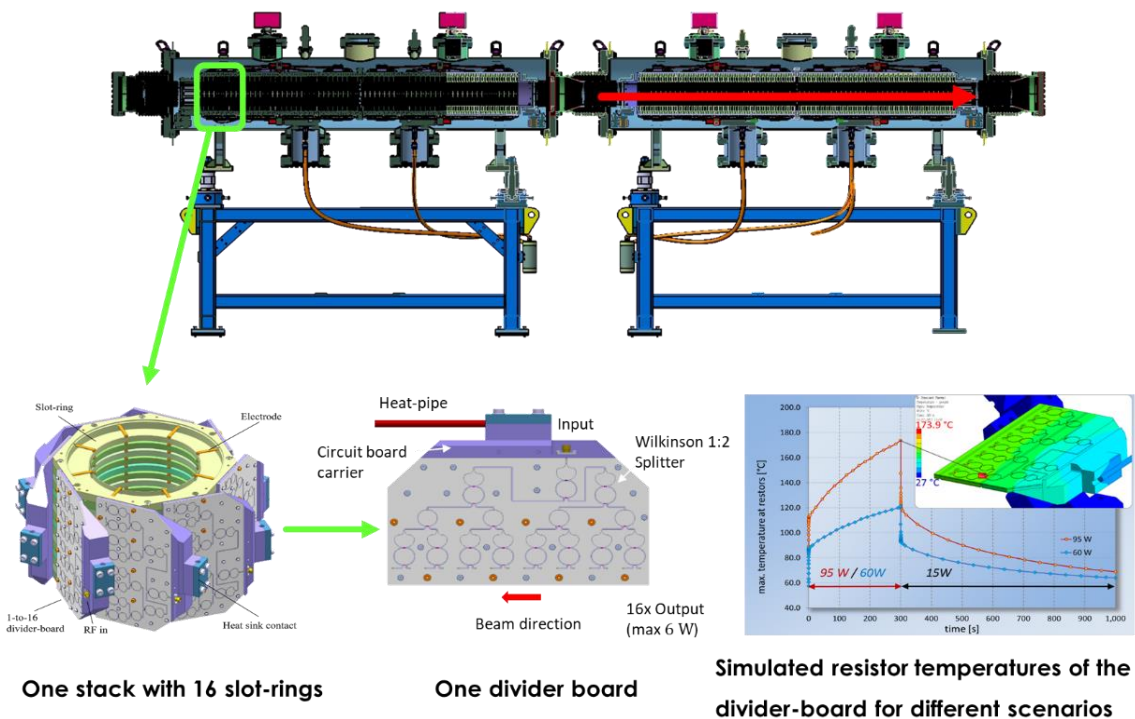

**Fig. 14.** Kickers designed by FZJ for the FAIR CR-SC System [18, 19].

As a first attempt, a simplified cooling plate (in a rectangular form measuring 48.5 mm × 97.0 mm × 10.0 mm) was designed and planned to be additively manufactured from pure copper, featuring a 4.0 mm diameter internal cooling channel shaped as the "SCO" (Stochastic Cooling Group) logo (see Fig. 15). Details of the divider board, including the actual board shape and resistor layout, will be incorporated in a subsequent design phase for the final version of the cooling plate.

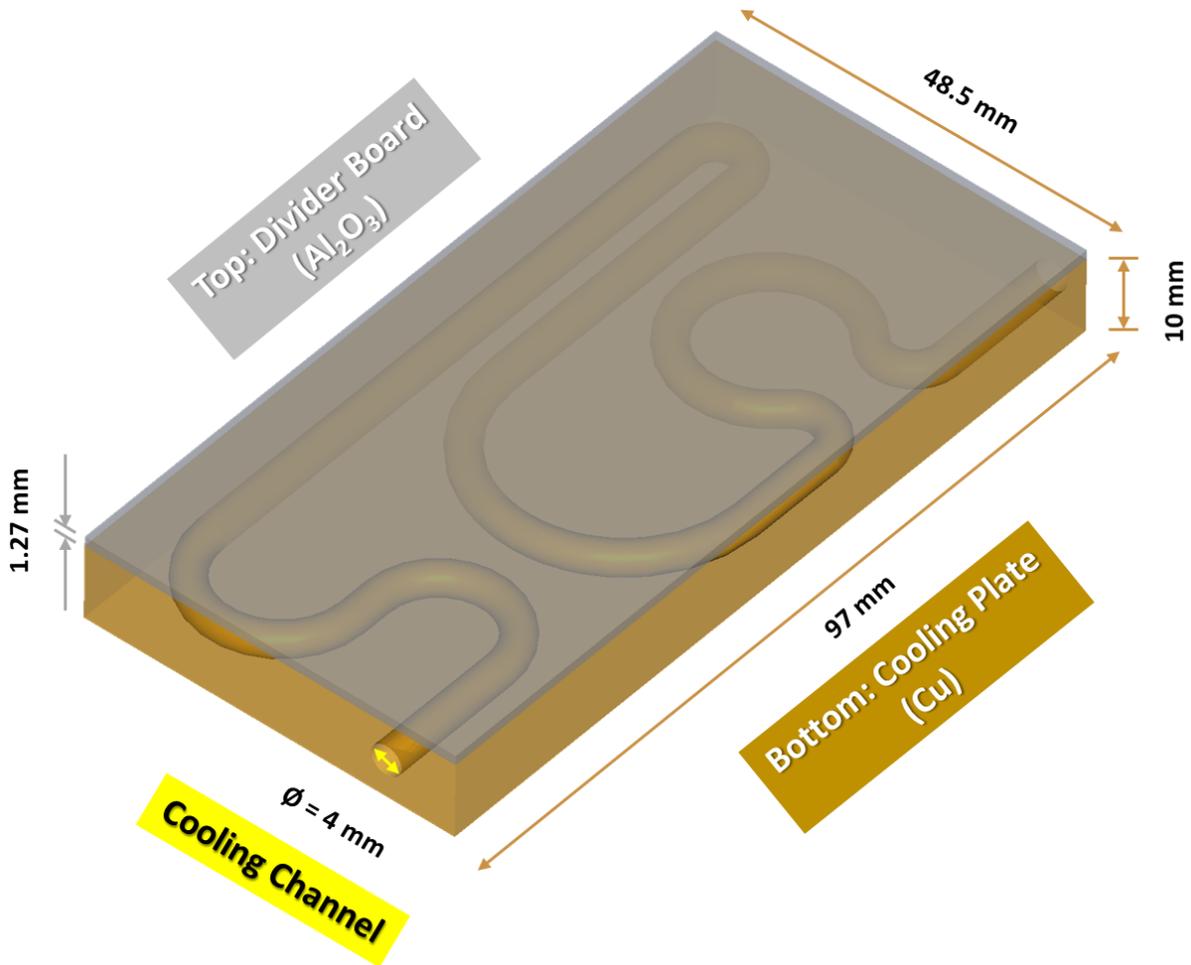

**Fig. 15.** The proposed cooling-plate prototype featuring a 4.0 mm diameter internal cooling channel shaped as the "SCO" logo (with the divider board illustrated in grey).

The settings adopted for the preliminary thermal simulations of the cooling-plate prototype can be found in Table 7. In the simulations, a heat load of 600 W (including a safety margin) was applied uniformly to the top surface of a 1.27 mm thick ceramic ($Al_2O_3$) layer representing the divider board (see Fig. 15). Figure 16 (a) shows that to keep the water velocity below 1.20 m/s, the maximum surface temperature on the divider board reaches 66.2°C, and the outlet water temperature approaches 45°C. To achieve a maximum surface temperature comparable to the CH case shown in Fig. 3 (a), specifically 42.2°C, a water velocity of approximately 4.4 m/s would be necessary. However, such a high flow rate is undesirable for copper due to the accelerated risk of metal erosion. To reduce the maximum surface temperature while maintaining an adequate operational lifetime, we proposed a new cooling-plate design that combines copper for the bulk body and stainless steel 316L for the inner cooling channels, fabricated via additive manufacturing. The reasoning was as follows:

- As listed in Table 6, stainless steel 316L and copper have similar thermal expansion coefficients, typically around $16 \times 10^{-6}$ /°C, which means they expand and contract at nearly the same rate when heated or cooled. This similarity helps minimize mechanical stress caused by differences in expansion when the two materials are combined.
- Stainless steel 316L is mechanically more robust than copper but exhibits a lower thermal conductivity. Thus, copper was selected for the bulk-body material to secure good heat transfer, while a thin stainless-steel wall (2 mm thickness) was applied to protect the cooling channel against water-induced damage.

Since the commercially available dual-material printing process does not offer pure copper, CuCr1Zr (a copper alloy with slightly lower thermal conductivity but higher hardness) was used instead. Corresponding ANSYS simulations were performed. As shown in Fig. 16 (b), the dual-material design results in a maximum surface temperature of 54.2°C with a maximum water velocity of 4.43 m/s, which is considered acceptable. The figure also reveals that the temperature distribution across the ceramic layer is not so uniform and hot areas are concentrated in zones beyond the coverage of the water channel. The current prototype uses a single cooling channel with a given shape; however, later the cooling channels in the final design will be matched to the actual heat load distribution of the ceramic divider board, which is expected to further reduce the peak surface temperature.

**Table 7.** Comparison of main design parameters and ANSYS Discovery simulation results of the single-material and dual-material SCO cooling plates.

| Parameter | Single-Material | Dual-Material |
|---|---|---|
| Materials | all: pure copper | body: CuCr1Zr<br>cooling channel: stainless steel 316L (wall thickness 2: mm) |
| Thermal load on the divider board [W] | 600 | 600 |
| Water pressure [bar] | 6 | 6 |
| Inlet water temperature $T_{water,i}$ [°C] | 25 | 25 |
| Inlet water flow rate $\dot{m}$ [kg/s] | 0.007 | 0.033 |
| Outlet water temperature $T_{water,o}$ [°C] | 44.9 | 29.2 |
| Max. water velocity $v_{water,max}$ [m/s] | 1.18 | 4.43 |
| Max. surface temperature on the divider board $T_{surface,max}$ [°C] | 66.2 | 54.2 |

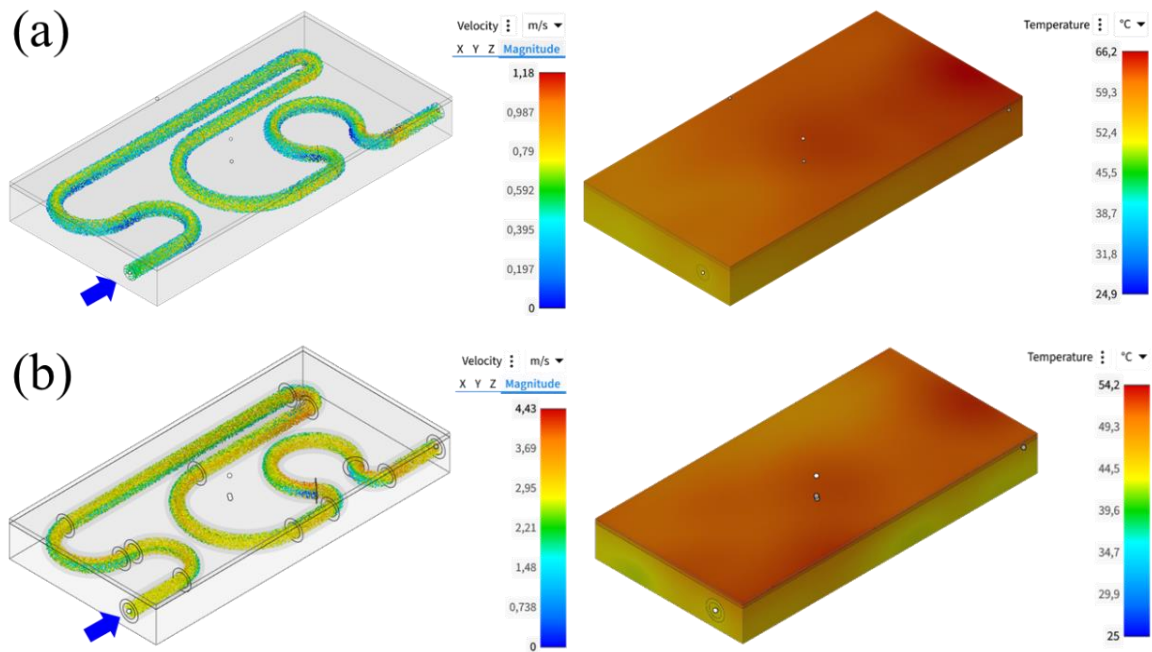

**Fig. 16.** Thermal simulations performed for the "SCO" cooling plates with different materials (top: pure copper; bottom: CuCr1Zr and stainless steel 316L).

The printed "SCO" prototype and its corresponding CT-scan image are shown in Fig. 17. The printing quality, in terms of both surface finish and internal cooling channel structure, is observed to be lower than that of the printed RACE cavity or the SHAPER cavity piece.

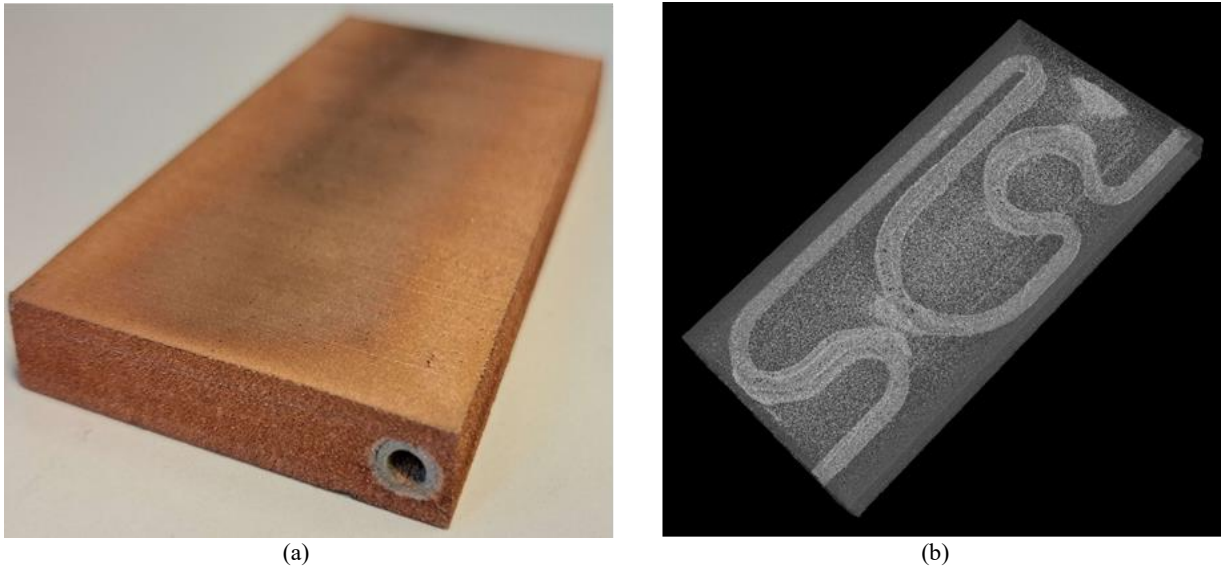

(a)                                                                 (b)

**Fig. 17.** (a) The dual-material-printed "SCO" prototype; (b) the CT-scan of the dual-material-printed "SCO" prototype.

One possible explanation is that printing with two metallic materials is inherently more complex than with a single material. Another contributing factor may be the substantial difference in thermal conductivity between CuCr1Zr and stainless steel 316L, despite their similar coefficients of thermal expansion. CuCr1Zr exhibits a high thermal conductivity of approximately 310 W/(m·K), whereas stainless steel 316L has a much lower value of about 15 W/(m·K). Consequently, CuCr1Zr can dissipate heat more rapidly, while stainless steel 316L is comparatively inefficient in heat conduction. Under thermal loading, the similar rates of expansion do not prevent the development of temperature gradients between the copper and stainless-steel regions. These gradients induce internal thermal stresses, particularly at the interface, which may in turn promote microcrack formation or fatigue over time.

Meanwhile, the feasibility of employing the high-performance alloy, CuNi2SiCr, as the sole printing material for the entire cooling plate has been also investigated. As mentioned earlier, CuNi2SiCr offers hardness comparable to stainless steel 316L while exhibiting significantly higher thermal conductivity. ANSYS simulations demonstrate that, under the same conditions as the dual-material scenario, $T_{\text{surface,max}}$ can be reduced to 48.2 °C with CuNi2SiCr. Moreover, adopting a single material is expected to improve the printing quality substantially.

We will continue the design and development of a practical cooling plate, taking into account the detailed structure of the divider board and incorporating water connections to facilitate realistic cooling tests.

## 4. Summary

An overview of ongoing activities applying additive manufacturing technologies to the R&D of particle accelerators and storage rings at GSI and for the FAIR project has been presented.

Each of these three projects —RACE, SHAPER, and SCO— is characterized by unique characteristics. In the RACE project, the highest-frequency CH cavity to date was realized using AM, including the invention of novel curved, slim cooling channels. In SHAPER, the first sextupole H-mode cavity has been developed to extend the H-mode DTL family, together with a special piecewise fabrication approach. The SCO project is the first application of additive manufacturing to stochastic cooling R&D, where dual-material printing has been tested.

Across these projects, we explored various cooling-channel designs, advanced printing materials—including new alloy materials for accelerators such as CuCr1Zr and CuNi2SiCr—as well as different fabrication techniques. Both successful experiences and lessons learned were gathered. One challenge was that the surface quality of printed components was not yet good enough for direct use. But this is not a limiting factor. An additional material allowance (e.g., 1 mm) can be included during printing and subsequently removed through careful post-processing to achieve the required dimensions precisely. Moreover, the post-processing is expected to become progressively simpler, because hybrid "print-and-process" systems are currently being developed in the industry. Also, measurements on printed components confirmed that vacuum requirements are achievable after post-processing. In addition, copper plating can be applied to ensure good RF performance.

To sum up, additive manufacturing is a highly promising technology for the development of future accelerator and storage-ring components. Additive manufacturing can not only deliver notable performance enhancements through its design freedom—such as curved, slim channels for improved cooling—but also significantly reduce construction costs and time. This is especially crucial for constructing large facilities with diverse resonant structures, as a single printer can, in principle, produce any required structure (as long as the structural dimensions are within the allowable range of the construction room of the printer).

Our future work will include:
- Completing the R&D of the three projects and testing the printed cavities and cooling plate with real operation.
- Further improving and developing more efficient cooling-channel designs and cavity geometries by leveraging the design freedom enabled by additive manufacturing.
- Conducting a comprehensive investigation of candidate construction materials for accelerators, including studying differences between bulk and printed materials, understanding anisotropy related to printing orientation, and evaluating / developing new materials for accelerator R&D.

We believe that additive manufacturing will revolutionize accelerator and storage-ring technologies, greatly expanding societal access to accelerator-based applications.

## 5. Acknowledgement

The authors would like to express their sincere gratitude to the colleagues at GSI Darmstadt—Prof. Dr. R. Aßmann, Dr. U. Weinrich, Dr. G. Walter, Dr. M. Heilmann, Dr. R. Gebel, A. Bardonner, N. Shurkhno, M. Römig, S. Teich, R. Erlenbach, M. Uhlig, N. Kähne, M. Henke, Dr. A. Schnase, Dr. A. Seibel, and Dr. C. Kleffner—, at IAP Frankfurt—Prof. Dr. H. Podlech, S. Wagner, J. Storch, and D. Bänsch—, and at FZ Jülich— Dr. R. Stassen— for their valuable support and help. Special thanks are extended to the GSI Mechanical Workshop and the GSI Technology Lab for their professional contributions. The authors also gratefully acknowledge N. Guillaume and F. Fanny from GANIL, France, for the kind discussions on the feasibility of a beam test of our printed cavity at SPIRAL2.